\documentclass[aps,prl,twocolumn,superscriptaddress,preprintnumbers,%
               showpacs,nofootinbib]{revtex4}
\newcommand{\PRE}[1]{}       % Use if journal style
\usepackage{bm}
\usepackage{epsfig}

\newcommand{\postscript}[2]{\setlength{\epsfxsize}{#2\hsize}
   \centerline{\epsfbox{#1}}}

\def\eslt{\not\!\!\!{E_T}}
\def\to{\rightarrow}

\def\bi{\begin{itemize}}
\def\ei{\end{itemize}}

\def\tb{\tilde b}

\def\tst{\tilde t}

\def\tg{\tilde g}

\def\tq{\tilde q}
\def\tw{\widetilde W}
\def\tz{\widetilde Z}

\def\alt{\lesssim}

\def\be{\begin{equation}}  
\def\ee{\end{equation}}  
\def\bea{\begin{eqnarray}}  
\def\eea{\end{eqnarray}}  
\def\delew{\Delta_{\rm EW}}

\begin{document}

\preprint{OUHEP-170130}
\preprint{UH-511-1275-17}

\title{
\PRE{\vspace*{1.5in}}
What hadron collider is required 
to discover or falsify natural supersymmetry?
%Testing Natural Supersymmetry:
%A Rationale for the High Energy LHC
%Which colliders are required for testing the\\
%weak scale supersymmetry hypothesis?
%The right hadron collider for testing\\ 
%weak scale supersymmetry
\PRE{\vspace*{0.3in}}
}

\author{Howard Baer}
\affiliation{Dept. of Physics and Astronomy,
University of Oklahoma, Norman, OK, 73019, USA
\PRE{\vspace*{.1in}}
}
\author{Vernon Barger}
\affiliation{Dept. of Physics,
University of Wisconsin, Madison, WI 53706, USA
\PRE{\vspace*{.1in}}
}
\author{James S. Gainer}
%\affiliation{Dept. of Psychics and Astrology,
\affiliation{Dept. of Physics and Astronomy,
University of Hawaii, Honolulu, HI 96822, USA
\PRE{\vspace*{.1in}}
}
\author{Peisi Huang}
\affiliation{Mitchell Institute for Fundamental Physics and Astronomy,
Texas A\& M University, College Station, TX 77843, USA
\PRE{\vspace*{.1in}}
}
\author{Michael Savoy}
\affiliation{Dept. of Physics and Astronomy,
University of Oklahoma, Norman, OK, 73019, USA
\PRE{\vspace*{.1in}}
}
\author{Hasan Serce}
\affiliation{Dept. of Physics and Astronomy,
University of Oklahoma, Norman, OK, 73019, USA
\PRE{\vspace*{.1in}}
}
\author{Xerxes Tata}
%\affiliation{Dept. of Psychics and Astrology,
\affiliation{Dept. of Physics and Astronomy,
University of Hawaii, Honolulu, HI 96822, USA
\PRE{\vspace*{.1in}}
}

%\date{January 1, 2017}

\begin{abstract}
\PRE{\vspace*{.1in}} 

Weak scale supersymmetry (SUSY) remains a compelling extension of the
Standard Model because it stabilizes the quantum corrections to the Higgs
and $W,\ Z$ boson masses. In natural SUSY models these corrections are, by
definition, never much larger than the corresponding masses. 
Natural SUSY models all have an {\it upper limit} on the gluino mass, too high
to lead to observable signals even at the high luminosity LHC.
However, in models with gaugino mass unification, the wino is
sufficiently light that supersymmetry discovery is possible in other
channels over the entire natural SUSY parameter space with no worse than 3\%
fine-tuning. Here, we examine the SUSY reach in more general models
with and without gaugino mass unification
(specifically, natural generalized mirage mediation), 
and show that the high energy LHC (HE-LHC),
a $pp$ collider with $\sqrt{s}=33$~TeV, will be able to detect the SUSY
signal over the entire allowed mass range. 
Thus, HE-LHC would either discover or conclusively falsify natural SUSY
with better than 3\% fine-tuning using a conservative measure that
allows for correlations among the model parameters.

\end{abstract}

\pacs{12.60.-i, 95.35.+d, 14.80.Ly, 11.30.Pb}
%12.60.-i   Models beyond the standard model
%95.35.+d   Dark matter

\maketitle

The discovery of a new scalar boson $h(125)$ at the CERN Large Hadron
Collider\cite{Aad:2012tfa} (LHC) has cemented the
Standard Model (SM) as the appropriate effective field theory
describing physics up to the weak scale $m_{\rm weak}\sim 200$
GeV.  However, in the SM, the quantum corrections to the Higgs boson
mass are quadratically sensitive to the scale of new physics
and exceed the observed
value of $m_h$ unless the cut-off scale, beyond which the SM
ceases to be a valid description,  is as low as $\Lambda \sim 1$ TeV.
As the cutoff $\Lambda$ grows beyond the TeV scale, increasingly
precise fine-tunings of SM parameters are required in order to
maintain $m_h$ at its measured value.
%Such fine-tunings in a field
%theory are generally symptomatic of some missing ingredient or other
%pathology within the effective theory.

It has long been known that extending the underlying
spacetime symmetry from the Poincar\'e group to the more general
super-Poincar\'e (supersymmetry or SUSY) 
group tames the quantum corrections to $m_h$, provided that SUSY
is softly broken not very far from the weak scale\cite{review}.
Realistic particle physics models incorporating SUSY,
such as the Minimal Supersymmetric Standard
Model (MSSM), thus require the existence of
new superpartners\cite{review2}, {\it some of whose masses lie
close to the weak scale},
%In realistic
%models, supersymmetry must be broken, albeit softly, so that
%cancellation of quadratic divergences is maintained. To stabilize
%$m_h$ without introducing destabilizing {\it log} divergences, it is
%generally expected that the new superpartners should exist not too far
%from the weak scale
hence the name {\it weak scale supersymmetry} (WSS); the
remaining ones may have multi-TeV masses.  Three
independent calculations involving virtual quantum effects provide
indirect experimental support for WSS. 1)~The measured
values of the three SM gauge couplings
unify at a scale $Q\simeq 2\times 10^{16}$ GeV in the MSSM
but not in the SM, 2)~the
top quark mass, $m_t\simeq 173$ GeV, falls within
the range required by SUSY to radiatively break
electroweak gauge symmetry, and 3)~the measured value of the Higgs
mass, $m_h\simeq 125$ GeV, (which could have taken on any value up
to the unitarity limit $\lesssim 1$ TeV in the SM) falls
within the narrow range, $m_h<135$
GeV\cite{Carena:2002es}, required by the MSSM. 

These considerations led many to expect WSS to be
discovered once sufficient data were accumulated at the LHC.
However, with nearly 40 fb$^{-1}$ of data at $\sqrt{s}=13$ TeV,
no evidence for superpartner production has been
reported.  Recent analyses based on $\sim 36$ fb$^{-1}$ of integrated
luminosity have produced mass limits on the gluino $\tg$ (spin$-1/2$
superpartner of the gluon) of $m_{\tg}>2$ TeV
and of the top squark (the lighter of the spin$-0$ superpartners of
the top quark) of $m_{\tst_1}>1$ TeV\cite{stop1} (within
the context of various simplified SUSY models), with even stronger limits on
first generation squarks. 
These may be compared with early estimates -- 
based upon the naturalness principle that 
{\it contributions to an observable} (such as the $Z$-boson mass) 
{\it should be less than or comparable to its measured value} -- that the
upper bound on $m_{\tg}$ is $\sim 350$ GeV and that $m_{\tst_1}\alt
350$ GeV based on
no less than $3\%$ fine-tuning\cite{Barbieri:1987fn}.\footnote{
%Lest the reader object that naturalness is a purely subjective quality,
We recall three cases where naturalness correctly presages 
the onset of new physics:
1. the classical electromagnetic contributions to the electron energy
$E=m_ec^2$ required a relativistic treatment of spacetime and its 
concommitant {\it positron}\cite{MUR}, 
2. the electromagnetic mass difference 
of the charged and neutral pions required new physics below $\sim 850$ MeV 
(matched by $m_{\rho}\simeq 770$ MeV)\cite{GG} and 
3. a computation of the $K_L-K_S$ mass difference required the 
existence of the charm quark with $m_c\sim 1-2$ GeV\cite{GL}. }
%These latter
%upper bounds have been challenged in that they were calculated within
%the context of multiple soft-parameter effective theories where the
%multiple soft parameters are introduced to parameterize one's
%ignorance as to the origin of SUSY breaking.
Similar calculations seemed to require {\it three} third generation squarks
lighter than 500 GeV\cite{Kitano:2006gv,Papucci:2011wy}.
%These
%calculations were challenged in that they neglected the soft Higgs
%self contribution $m_{H_u}^2$ contribution to renormalization group
%running.
Crucially, the analyses leading to these stringent upper bounds assume
that contributions to the radiative corrections from various
superpartner loops are {\it independent}. 
The assumption of independent soft terms is not valid in
frameworks where the seemingly independent parameters -- introduced to
parametrize our ignorance of the underlying SUSY breaking dynamics -- 
are in fact correlated as in a more fundamental theory\cite{bbm,AM,Baer:2014ica}.
It has been argued that ignoring these
correlations leads to prematurely discarding viable SUSY
models; allowing for such correlations leads to
the possibility of {\it radiatively-driven
naturalness}\cite{Baer:2012up,Baer:2012cf} where large, 
seemingly unnatural values of GUT scale soft terms (such as $m_{H_u}^2$) 
can be radiatively driven to
natural values at the weak scale due to the large value of the
top-quark Yukawa coupling.

%Allowing for radiatively driven naturalness, then

Indeed, it has been shown that to allow for the possibility of parameter
correlations one should only require that
the {\it weak scale}
contributions to $m_Z$ (or $m_h$) be not much larger than
their measured values. From minimization of the MSSM scalar potential,
one can relate $m_Z$ to weak scale MSSM Lagrangian parameters
\bea
\frac{m_Z^2}{2}&=&\frac{m_{H_d}^2+\Sigma_d^d-(m_{H_u}^2+\Sigma_u^u)\tan^2\beta}{\tan^2\beta -1}-\mu^2. \label{eq:mzs1}
\eea
%&\simeq& -m_{H_u}^2-\Sigma_u^u-\mu^2,
%\label{eq:mzs}
%\eea
%
Here $\Sigma_u^u$ and $\Sigma_d^d$ denote 1-loop corrections
(expressions can be found in the Appendix of Ref. \cite{Baer:2012cf}) to the
scalar potential, $m_{H_u}^2$ and $m_{H_d}^2$  the Higgs soft masses
at the weak scale,
and $\tan \beta \equiv \langle H_u \rangle / \langle H_d \rangle$.
%The second line is obtained for moderate to large values of $\tan\beta
%\agt 5$ (as required by the Higgs mass calculation\cite{Carena:2002es}).
SUSY models requiring large cancellations between the various terms on the
right-hand-side of Eq.~(\ref{eq:mzs1}) to reproduce the measured value of
$m_Z^2$ are regarded as unnatural, or fine-tuned.
%In contrast, SUSY
%models which generate terms on the RHS of Eq.~\ref{eq:mzs} which are
%all less than or comparable to $m_{\rm weak}$ are regarded as natural. 
Thus, natural SUSY models are characterized by low values of the {\it
  electroweak} naturalness measure $\delew$ defined as
\cite{Baer:2012up,Baer:2012cf}.
%(for some recent discussion, see \cite{Ross:2017kjc})
%
\be \delew\equiv \text{max}|{\rm each\ term\ on\ RHS\ of\
Eq.}~\ref{eq:mzs1}|/(m_Z^2/2).  \ee 

Since $\delew$, by definition, does not include large logarithms of the
high scale $\Lambda$, $\delew$ is smaller than the traditional
fine-tuning measures $\Delta_{\rm BG}$ \cite{Barbieri:1987fn} or
$\Delta_{\rm HS}$ \cite{Kitano:2006gv,Papucci:2011wy}. These logarithms
essentially cancel if the underlying model parameters are appropriately
correlated, and then the traditionally used fine-tuning measure reduces
to $\delew$ once these correlations are properly implemented
\cite{bbm,AM,Baer:2014ica}. We conservatively advocate using $\delew$
for discussions of fine-tuning since this automatically allows for the
possibility that underlying SUSY breaking parameters might well be
correlated. Disregarding this may lead to prematurely discarding perfectly
viable theories because the traditional computation of fine-tuning
(ignoring possible parameter correlations) may falsely lead us to
conclude that the model is unnatural.

%It has been shown that the high scale measures of fine-tuning
%$\Delta_{\rm BG}$\cite{Barbieri:1987fn} and 
%$\Delta_{\rm HS}$\cite{Kitano:2006gv,Papucci:2011wy} reduce to $\delew$
%once underlying correlations between parameters are properly
%incorporated\cite{AM,bbm,Baer:2014ica}. 
%Furthermore, $\delew$ is the most conservative of these measures
%in that it yields the correct measure of fine-tuning allowing 
%for the fact that various soft SUSY breaking parameters might be 
%correlated in the underlying theory. 
%As a result,  $\delew<\Delta_{\rm BG}<\Delta_{\rm HS}$\cite{bbm} so 
%that if a model is fine-tuned under $\delew$, then it appears even more 
%fine-tuned under $\Delta_{\rm BG}$ or $\Delta_{\rm HS}$. 
%$\delew<\Delta_{\rm BG}<\Delta_{\rm HS}$\cite{bbm} so 
%that if a model is fine-tuned under $\delew$, then it appears even more
%fine-tuned under $\Delta_{\rm BG}$ or $\Delta_{\rm HS}$.
%(Also, $\delew$ is a model-independent measure in that all models 
%yielding the same sparticle mass spectrum will have the same value of 
%$\delew$;
%this property is not shared by $\Delta_{\rm BG}$ or $\Delta_{\rm HS}$.)

We see from Eq.~(\ref{eq:mzs1}) that the robust
criteria for naturalness are the weak scale values:
\bi
\item $m_{H_u}^2\sim -(100-300)^2$ GeV$^2$, and
\item $\mu^2\sim (100-300)^2$ GeV$^2$\cite{Chan:1997bi}
  \ei
(the lower the better).
For moderate-to-large $\tan\beta \gtrsim 5$, the remaining contributions
other than $\Sigma_u^u$ are suppressed.  The largest radiative
corrections $\Sigma_u^u$ typically come from the top squark
sector. The value of the trilinear coupling $A_0 \sim -1.6m_0$ leads to
split TeV-scale top squarks and minimizes $\Sigma_u^u(\tst_{1,2})$,
simultaneously lifting the Higgs mass $m_h$ to $\sim 125$~GeV
\cite{Baer:2012cf}. 

\begin{figure}[tbp]
\postscript{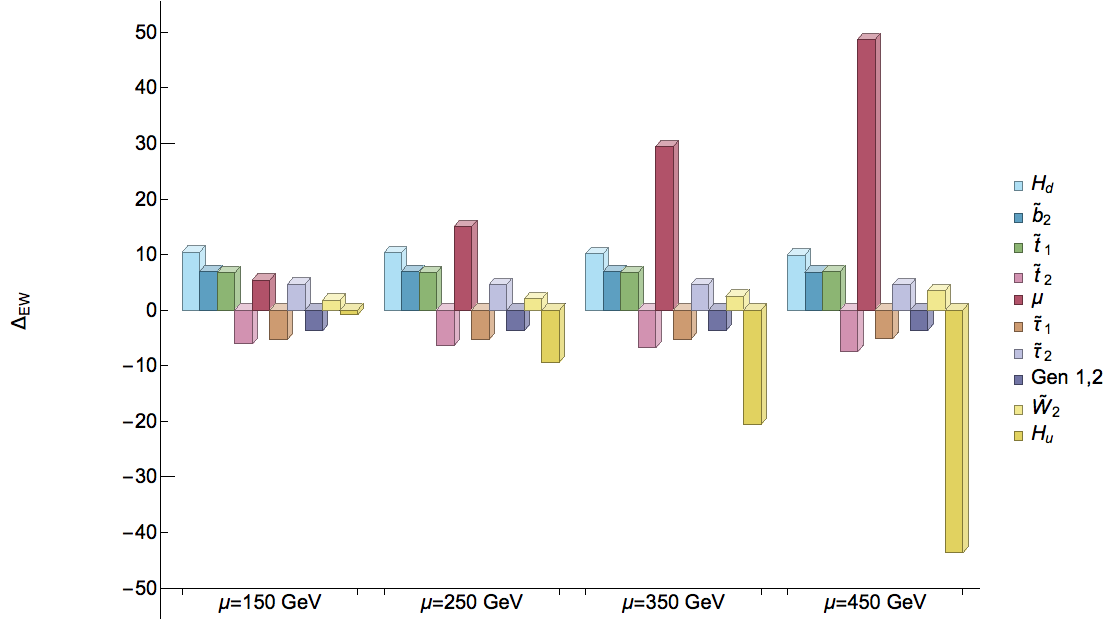}{1.0}
\caption{Top ten contributions to $\delew$ from NUHM2 model benchmark points 
with $\mu =150$, 250, 350 and 450 GeV.
\label{fig:dew}}
\end{figure}

A visual display of the top ten contributions to $\delew$ is shown in
Fig.~\ref{fig:dew} for NUHM2 benchmark points with $\mu =150$, 250, 350
and 450 GeV.  For $\mu =150$ GeV, all contributions to $m_Z$ -- some
positive and some negative -- are comparable to or less than the
measured value so the model is very natural.  For $\mu =250$ GeV with
$\delew=15$, we see that some fine-tuning is on the verge of setting in
so that the value of $m_{H_u}^2(weak)$ must be adjusted to compensate
for such a large value of $\mu$.  By the time $\delew \sim 30$,
corresponding to $\mu\sim 350$ GeV, cancellation between
(presumably) unrelated large contributions is clearly required. 
This value will therefore serve as a rather conservative upper limit on  $\delew$ 
in our study, since--
as we are considering ``natural SUSY''--
we expect the contributions to any observable (in this case $m_Z$) 
to be comparable to or less than the value of the observable.\footnote{
For concreteness we must choose some upper bound on $\delew$, 
and there is inherently subjectivity in this choice.
Since $\mu \gtrsim 100$ GeV (from LEP2 chargino search limits), then 
$\delew$ is necessarily $>1$, 
while it would be hard to describe $\delew \gtrsim 100$ as ``natural''.
The value $\delew =30$ corresponds to individual contributions to the 
right-hand-side of Eq. 1 which exceed a factor of $\gtrsim (3 m_Z)^2$.}
To obtain upper bounds on sparticle masses from naturalness, 
we therefore require $\delew <30$ (no worse than 3\% fine-tuning, 
even allowing for the fact that model parameters may be correlated).

\begin{figure}[tbp]
\postscript{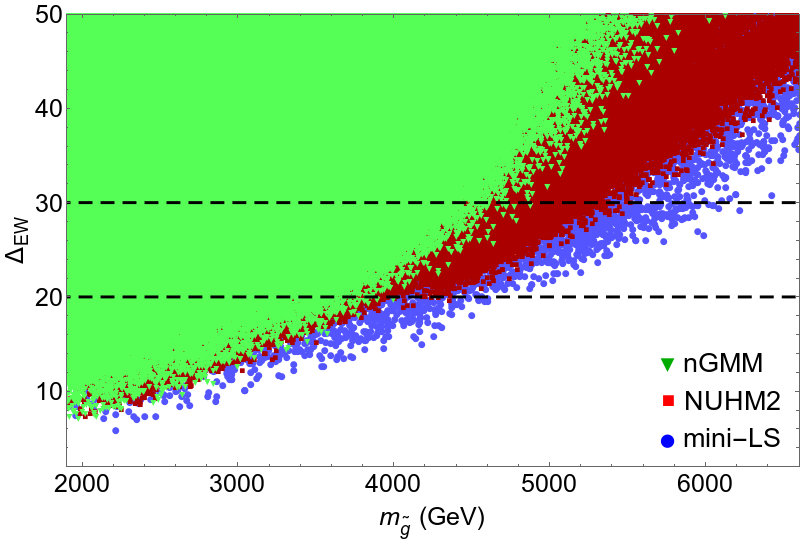}{1.0}
\caption{Plot of $m_{\tg}$ vs. $\delew$ from scan over NUHM2 model (red
squares), nGMM model (green traingles) and the mini-LS picture (blue
circles).  Points with $\delew < 30$ are conservatively regarded as
natural.
\label{fig:nGMM}}
\end{figure}

A large assortment of popular SUSY models with $m_h\simeq 125$ GeV were
examined in Ref. \cite{Baer:2014ica} where only the two-extra-parameter
(compared to the well-known mSUGRA/CMSSM model) non-universal Higgs
model (NUHM2)\cite{nuhm2} (with the two extra parameters $\mu$ and $m_A$
allowed to be free) was found to allow for naturalness.
Requiring $\delew<30$ in the NUHM2 model, then it was found 
that\cite{Baer:2012cf,Baer:2015rja} 
\bi
\item $m_{\tg}\alt 5$~TeV (see also Fig.~\ref{fig:nGMM}),
\item $m_{\tst_1}\alt 3$ TeV (with other third generation squarks 
bounded by $\sim 8$~TeV) and
\item $m_{\tw_1,\tz_{1,2}}\alt 300$ GeV,
\ei 

while other sfermions could be in the multi-TeV
range.  Thus, gluinos and squarks may easily lie beyond the reach of LHC
at little cost to naturalness with only the higgsino-like lighter
charginos and neutralinos required to lie close to the weak scale.\footnote{
Our conclusion about the existence of light higgsinos arises
from the fact that the higgsino mass is given by the superpotential 
parameter $\mu$ and this same parameter enters the Higgs boson mass calculation.
This situation can be circumvented in extended SUSY models with 
additional weak scale superfields beyond
those of the MSSM that have extra symmetries
\cite{Nelson:2015cea,Cohen:2015ala,Martin:2015eca} or in models where
SUSY breaking higgsino terms are allowed \cite{Ross:2017kjc}. If these
higgsinos couple to SM singlets, such terms would lead to a hard
breaking of SUSY.} 
The lightest higgsino $\tz_1$ comprises a portion of the dark matter and
would escape detection at LHC.  The remaining dark matter abundance
might be comprised of, {\it e.g.}, axions\cite{Bae:2013bva}.  Owing to
the compressed spectrum with mass gaps $m_{\tw_1}-m_{\tz_1}\sim
m_{\tz_2}-m_{\tz_1}\sim 10$--20~GeV, the heavier higgsinos are difficult
to see at LHC because the visible energy released from their decays
$\tw_1\to f\bar{f}'\tz_1$ and $\tz_2\to f\bar{f}\tz_1$ (where the $f$
denotes SM fermions) is very small.  The NUHM2 model can be embedded in
a general $SO(10)$ SUSY GUT.

Keeping in mind that the stabilization of the Higgs sector remains a
key motivation for WSS, these
upper bounds are vital for testing the validity of
the naturalness hypothesis.\footnote{We stress that WSS
  always resolves the big gauge hierarchy problem; we are
  concerned here with stabilizing the weak scale
  without the need for part per mille fine-tuning.}
While the naturalness upper bound is $m_{\tg}\alt 5$ TeV,
experiments at the LHC have probed $m_{\tg}< 1.9$~TeV
via the $\tg\tg$ production channel. 
The reach of the high luminosity LHC (HL-LHC) for gluino pair production 
has recently been evaluated in Ref. \cite{Baer:2016wkz} 
(see also \cite{atlas_reach} and \cite{CMS:2013xfa}). 
Using hard $\eslt$ cuts, it
was found that the LHC14 reach extends to $m_{\tg}\sim 2.4$ (2.8)~TeV for
300 (3000) fb$^{-1}$-- not sufficient to probe
the entire natural SUSY range of gluino masses.\footnote{Thus, Ref. \cite{Baer:2016wkz} and this paper answer the question posed in the Abstract to Ref. \cite{shafi}.}
Moreover, the HL-LHC is expected to probe
maximally to $m_{\tst_1}\sim 1.4$ TeV\cite{atlas_reach,CMS:2013xfa}, again
far short of the complete range of natural models.

This is not the complete story for the NUHM2 framework, because
the underlying assumption of gaugino mass unification
constrains the wino mass to be $\sim m_{\tg}/3$.  As
LHC integrated luminosity increases, wino pair production provides a
deeper reach into parameter space, via the clean same-sign diboson (SSdB)
channel\cite{ssdb} (from $pp\to\tw_2^\pm \tz_4$ with
$\tw_2^\pm \to W^\pm\tz_{1,2}$ and $\tz_4\to W^\pm\tw_1^\mp$).
This channel offers a
HL-LHC 3000 fb$^{-1}$ reach to $m_{1/2}\sim 1.2$ TeV, covering
nearly all of the $\delew <30$ region. Although electroweak production of
higgsinos is swamped by SM backgrounds due to the small visible
energy release in higgsino decays, higgsino pair production in
association with a hard QCD jet-- for instance $pp\to\tz_1\tz_2+jet$
with $\tz_2\to\tz_1\ell^+\ell^-$-- 
offers a HL-LHC reach to $\mu\sim 250$ GeV\cite{llj}. 
The presence of the soft dilepton pair with $m_{\ell\ell} <
m_{\tz_2}-m_{\tz_1}$ is crucial for limiting the SM
background. In general models (see below), where the wino is heavier than its
unification value, the SSdB signal would be kinematically suppressed,
and at the same time, the mass gap between the higgsinos would be
reduced, leading to a diminished efficiency for detection of the soft
leptons in the $\ell^+\ell^-+$monojet events just discussed. 
Thus although these combined channels cover nearly all of $\delew <30$ 
parameter space in the NUHM2 model or in the other low $|\mu|$ models 
with gaugino mass unification\cite{Baer:2016usl}, 
they cannot be relied on to guarantee
LHC detection in a natural SUSY framework without gaugino mass unification.

This leads us to examine the natural SUSY parameter space of an
alternative framework dubbed natural Generalized Mirage Mediation (nGMM)
in which the weak scale gaugino masses have (nearly) comparable
values. GMM is a generalization of well-motivated mirage mediation (MM)
models\cite{mirage} that emerge from string theory, with moduli fields
stabilized via flux compactification. Gaugino mass unification at the
{\it mirage unification scale} $\mu_{\rm mir}$, is the robust
characteristic of this scenario and leads to nearly degenerate
gauginos at the weak scale
if $\mu_{\rm mir}$ is close to $m_{\rm weak}$. Although MM
models that are based on simple compactification schemes appear to be
unnatural for the observed value of $m_h$\cite{Baer:2014ica}, 
a more general construction\cite{Baer:2016hfa} which allows for 
more diverse scalar soft terms allows
$\Delta_{\rm EW}< 30$ with $m_h=125$~GeV without altering the predicted gaugino
mass pattern. Thus nGMM models with low values of $\mu_{\rm mir}$ and
$m_{\tg}=3-4.8$~TeV may have very heavy winos, suppressing the SSdB
signal and leading to very small higgsino mass gaps (2-6~GeV) making
the $\ell^+\ell^-j+\eslt$ signal challenging to detect.
We see that the nGMM model presents a natural,
well-motivated framework which may well be beyond the HL-LHC reach.

The string-inspired natural mini-landscape (mini-LS) \cite{nilles}
models, whose phenomenology was recently examined in Ref.~\cite{mini}, 
is yet another well-motivated example where the spectrum satisfies
electroweak naturalness but may not be accessible at the HL-LHC. 
The mini-LS scenario is closely related to the nGMM model in that gaugino
masses maintain the relations of mirage unification-- but it 
differs in that the first/second generation scalar mass soft parameters
are significantly larger than those of the third generation and Higgs sector.
Models with deflected mirage mediation\cite{Barger:2015xqk}, 
or models in which the field that breaks supersymmetry transforms 
as the {\bf 75} rep. of $SU(5)$\cite{Amundson:1996nw}
also lead to a compressed gaugino spectrum which may likewise
lie beyond the HL-LHC reach.

To assess the capability of testing SUSY naturalness in a relatively
model-independent way, we should not rely on signals which are
contingent upon the lightness of the wino relative to the gluino.  We
have therefore programmed the nGMM model into the Isasugra/Isajet 7.86
spectrum generator\cite{Paige:2003mg} (for details on parameter space,
see Ref. \cite{Baer:2016hfa}). 
This also allows us also to generate the mini-LS spectrum. 
Next, we have performed detailed scans over the
allowed parameter space, requiring $m_{\tg}>1.9$ TeV and $m_h:123-127$
GeV (allowing for $\pm 2$~GeV theory error in the Isasugra calculation
of $m_h$).  We show in Fig. \ref{fig:nGMM} a scatter plot of $\delew$
versus $m_{\tg}$ for both the nGMM model (green triangles), the NUHM2
model (red squares) and the mini-LS picture (blue circles). From the
plot, we read off an upper bound $m_{\tg}\alt 4.6\ (5.6) [6.0]$~TeV if
$\delew < 30$ in the nGMM (NUHM2) [mini-LS] model.  The bound is only
mildly sensitive to the specific assumption about high scale wino and
bino masses, but does depend on the hierarchy between first/second
generation scalar and the top squark masses.  Henceforth we regard the more
conservative $m_{\tg}<6.0$ TeV as representative of an upper limit on
$m_{\tg}$ in all natural SUSY models and explore prospects for {\em
gluino detection} at a variety of hadron colliders with a view to either
detecting or excluding supersymmetry with $\leq 3$\% electroweak
fine-tuning.

\begin{figure}[tbp]
\postscript{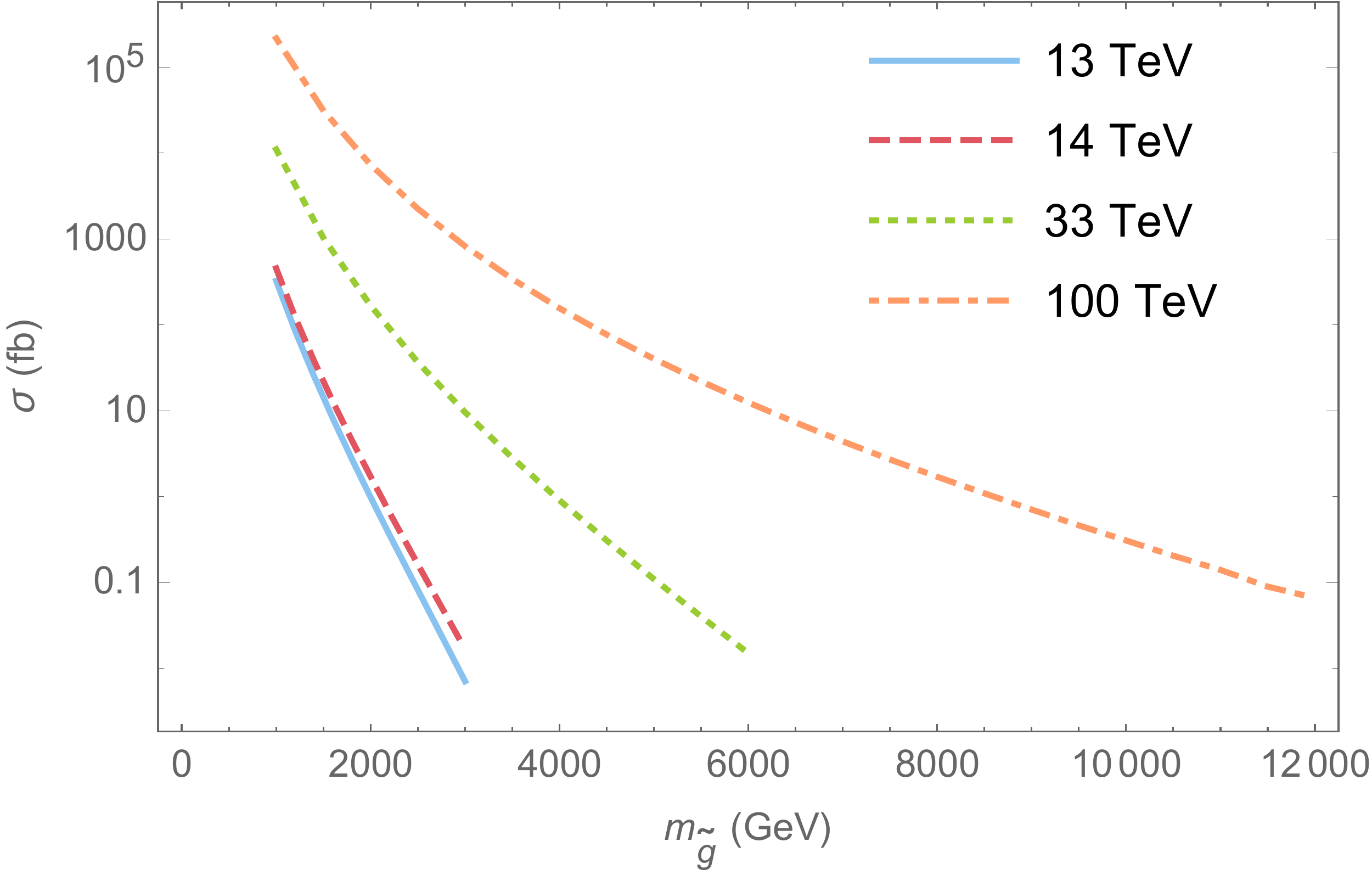}{1.0}
\caption{Total cross section (NLL+NLO) for gluino pair production
at various hadron colliders vs. $m_{\tg}$ for $m_{\tq} \gg m_{\tg}$.
\label{fig:xsec}}
\end{figure}

In Fig. \ref{fig:xsec}, we show the NLL+NLO evaluation\cite{Beenakker:1996ch} of $\sigma
(pp\to\tg\tg X)$ versus $m_{\tg}$ for $pp$ collider energies
$\sqrt{s}=13,\ 14,\ 33$ and $100$ TeV.  For 3000 fb$^{-1}$ at LHC14,
the gluino reach for the NUHM2 model extends out to $m_{\tg}\sim 2.8$
TeV\cite{Baer:2016wkz}, insufficient to probe the entire natural SUSY
parameter space in this channel. Naive scaling suggests that the
gluino reach would cover the entire natural SUSY range even at the HE-LHC,
a 33~TeV $pp$ collider, for which a peak luminosity of $2\times10^{34}$
~cm$^{-2}$ ~s$^{-1}$, corresponding to about 100~fb$^{-1}$ per operating year,
has been projected\cite{bruenning}.

Here, we perform a careful analysis of the natural SUSY reach via
gluino pair production at the HE-LHC, assuming the gluinos
primarily decay to third generation squarks as expected in natural
SUSY models. 
We have explored the reach in various multijet plus $\eslt$
channels and found that the greatest reach (as measured by statistical
significance of the signal over SM backgrounds) is obtained in the
$\ge 4j + \eslt$ channel with $\ge 2$ tagged $b$-jets. We use the same
$b$-jet tagging algorithm as in Ref.\cite{Baer:2016wkz} and find that
the reach is nearly optimized with the same set of cuts as in that
study, except that we now require jets to have $E_T> 200$~GeV and
require $\eslt > 1500$~GeV for the heavier gluinos under consideration. 

We perform our analysis for several model lines designed to capture
features of gluino events in natural SUSY models. We first examine
an NUHM2 model line with $m_0=5m_{1/2}$, $A_0=-1.6m_0$,
$m_A=m_{1/2}$, $\tan\beta=10$ and $\mu=150$~GeV. For this model line,
over the mass range of interest (2-6~TeV), the gluino always decays
via $\tg\to \tst_1 t$, with
$\tst_1\to b\tw_1$ at 50\%, $\tst_1\to t\tz_1$ at $\sim 25\%$ and
$\tst_1\to t\tz_2$ at $\sim 25\%$\cite{Baer:2016bwh}. 
The decay products of the daughter
higgsinos are essentially invisible. Gluino pair production gives rise
to final states with $tttt$, $tttb$ or $ttbb$ plus large $\eslt$. For
this model line $m_{\tst_1}$ increases with gluino mass and is
0.8-1~TeV below $m_{\tg}$ for $m_{\tg}=2-5$~TeV.  Since
the efficiency for detection after cuts will be
sensitive to event kinematics, we have also examined three 
simplified model lines with $m_{\tst_1}=1,2$ and 3~TeV independent of
$m_{\tg}$, where we assume the gluino always decays via $\tg\to t\tst_1$
and that the stop decays as in model line 1. We expect that these model
lines capture much of the variation expected from natural
SUSY models, including the possibility that some fraction of models
have a significant (but subdominant) branching fraction for gluino
decays to $\tst_2$ or $\tb_1$ squarks whose decays also lead to third
generation squarks in the final state. We have checked that
for most models with $\delew < 30$, $B(\tg\to \tst_1 t) \ge 60\%$.

\begin{figure}[tbp]
\postscript{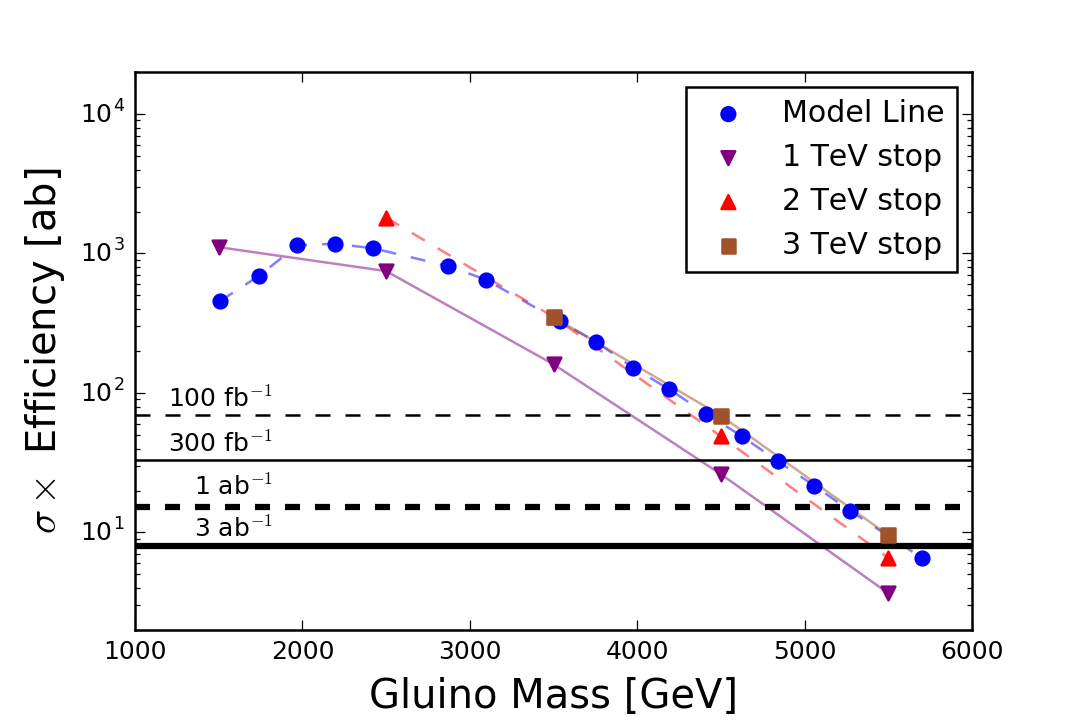}{1.0}
\caption{Plot of cross section after cuts in the $2$-tagged $b$-jet
  analysis along with $5\sigma$ discovery lines for 100, 300, 1000 and
  3000 fb$^{-1}$ for the NUHM2 model line introduced above (blue circles),
    as well as 
    simplified models with $m_{\tst_1}=1$~TeV (upside-down purple triangle),
    2~TeV (red triangles) and 3~TeV (brown squares).
\label{fig:2b}}
\end{figure}

The results of our computation of gluino signal cross section after
analysis cuts in the multijet plus $\eslt$ channel with $\ge 2$ tagged
$b$-jets is shown in Fig.~\ref{fig:2b} for the NUHM2 model line
introduced above (blue circles), as well as for the simplified models
with $m_{\tst_1}=1$~TeV (upside-down triangle), 2~TeV (triangle) and
3~TeV (squares).  We have checked that the cross section for a
simplified model line with $m_{\tst_1}=4$~TeV (and large enough gluino
masses) is very close to that for the first model line. The horizontal
lines denote the cross section levels required for a 5$\sigma$ signal
significance above SM backgrounds from $t\bar{t}$, $t\bar{t}t\bar{t}$,
$t\bar{t}b\bar{b}$, $Wt\bar{t}$, $Zb\bar{b}$ and single top
production.\footnote{If the background is underestimated/overestimated
by factor $f$, these horizontal lines will shift up/down, by about a
factor $\alt \sqrt{f}$. 
For $f =  2$ the reach projection is affected by only 
$\approx 100 - 150$~GeV for ab$^{-1}$ scale integrated luminosities. 
The effects of event pile-up depend on details of both machine and detector 
performance and thus are beyond the scope of the present analysis.
A discussion of pile-up for CMS at LHC14 is given in Ref. \cite{CMSPU}.} 
We see that, with an integrated luminosity of
1~ab$^{-1}$, the $5\sigma$ gluino mass reach at the 33~TeV machine
extends to $m_{\tg}= 4.8$~TeV (and covers the entire $\Delta_{\rm EW} <
20$ part of the allowed mass range) {\it even with the most pessimistic
assumption} for the top squark mass.\footnote{Our LHC33 reach values are
comparable to those values previously calculated for hadronic channels
in the context of simplified models in Ref's
\cite{Cohen:2013zla,Gershtein:2013iqa}.}

It should be kept in mind that this is an extremely conservative
estimate of the reach: a 1~TeV stop is just above the current bound, so such
scenarios will either be excluded or discovered well before HE-LHC
accumulates 1~ab$^{-1}$ of data.  We have also checked\cite{mini} that
in these natural SUSY models, $m_{\tg} > 4.8$~TeV only if $m_{\tst_1}<
2$~TeV, and further that the LHC33 reach for top squark comfortably
exceeds 2.7~TeV, assuming that the top squark dominantly decays to
higgsinos via $\tst_1 \to t\tz_1$, $\tst_1\to \tz_2$ and $\tst_1\to
b\tw_1$ with branching ratios 1:1:2 \cite{Baer:2017pba}. 
It is, therefore,
reasonable to conclude that a 33~TeV $pp$ collider will decisively probe
almost the entire range of gluino masses available to natural SUSY
models with no worse than 3\% electroweak fine-tuning, and that if the
gluino is too heavy for detection, the signal from the top squark will
definitely be accessible.

\begin{figure}[tbp]
\postscript{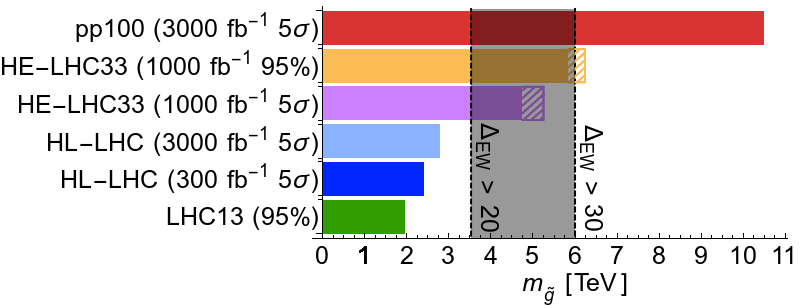}{1.0}
\caption{Reach of various hadron collider options for natural SUSY in
the gluino pair production channel compared to upper bounds on $m_{\tg}$
(gray band) in natural SUSY models. 
The hatches reflect some model dependence of the HE-LHC reach where
the lower edge is {\it very conservative} since the light stops
(for which the lower edge is calculated) offer an independent SUSY 
discovery channel\cite{Gershtein:2013iqa}.
\label{fig:bar}}
\end{figure}

In Fig. \ref{fig:bar}, the bars show several $5\sigma$ gluino discovery 
and 95\%CL exclusion reaches in natural SUSY
models for various $pp$ collider options via the
channel $pp\to \tg\tg$ along with the naturalness upper bound on $m_{\tg}$.
We expect that this upper bound is insensitive to the details of the
model as a pMSSM scan with $\delew < 30$ also yields the same bound
\cite{Baer:2015rja}. 
The region below the gray band is considered not fine-tuned while the 
region beyond is fine-tuned. 
We see that the HE-LHC discovery reach with $\sqrt{s}\sim 33$ TeV 
and 1000 fb$^{-1}$ will just about cover the entire natural SUSY 
parameter space as conservatively 
defined by $\delew < 30$. Moreover, if the gluino is too heavy to be
discovered, the top squark signal will be accessible.
Thus, HE-LHC should suffice to either discover or falsify natural
supersymmetry.  We also show the reach of a proposed $\sqrt{s}=100$ TeV
$pp$ collider (the FCC-hh or $SppC$) within the context of a simplified
model assuming gluino three-body decay to massless
quarks\cite{Gershtein:2013iqa}.  The 100 TeV $pp$ collider can probe to
values of $m_{\tg}$ over 10 TeV. (This is likely 
a conservative value since the projected reach would
likely extend to somewhat larger values if instead 
gluinos are assumed to dominantly decayed to third generation squarks.)
However, we note that HE-LHC should
already be able to discover or falsify natural SUSY within the context
of the MSSM at a small fraction of the cost of a 100 TeV machine.

In summary, supersymmetric models with weak scale naturalness are
well-motivated SM extensions with impressive indirect
support from measurements of gauge couplings and the top-quark and
Higgs boson mass.
While the HL-LHC appears  sufficient to probe natural
SUSY models with gaugino mass unification, 
we have shown that HE-LHC with $\sqrt{s}=33$~TeV is required
to either discover or falsify natural SUSY (with $\delew < 30$)
even in very general -- but equally natural -- SUSY scenarios such 
as nGMM with a compressed gaugino spectrum.
Alternatively, an $e^+e^-$ collider with $\sqrt{s}\sim 0.5-0.7$ TeV 
would be sufficient to either discover or falsify natural SUSY 
via pair production of the required light higgsinos\cite{Baer:2014yta}.
Discovery of natural SUSY via either of these machines 
would then provide enormous impetus for the construction of
even higher energy machines which could then access many of the
remaining superpartners. 

{\it Acknowledgements:} 
This work was supported in part by the US Department of Energy, Office of High
Energy Physics.

%%%%%%%%%%%%%%%%%%%%%%%%%%%%%%%%%%%%%%%%%%%%%%%%%%%%%%

%
\end{document}